# Gate reflectometry for probing charge and spin states in linear Si MOS split-gate arrays

L. Hutin[1], B. Bertrand[1], E. Chanrion[2], H. Bohuslavskyi[4], F. Ansaloni[4], T.-Y. Yang[5], J. Michniewicz[6],
D. J. Niegemann[2], C. Spence[2], T. Lundberg[5,6], A. Chatterjee[4], A. Crippa[3], J. Li[3], R. Maurand[3], X. Jehl[3], M. Sanquer[3],
M. F. Gonzalez-Zalba[5], F. Kuemmeth[4], Y.-M. Niquet[3], S. De Franceschi[3], M. Urdampilleta[2], T. Meunier[2], M. Vinet[1]

[1]CEA-Leti Minatec, Grenoble, France, email: louis.hutin@cea.fr, [2]CNRS Institut Néel, Grenoble, France,
[3]CEA-IRIG, Grenoble, France, [4]Niels Bohr Institutet, University of Copenhagen, Denmark,
[5]Hitachi Cambridge Laboratory, United Kingdom, [6]Cavendish Laboratory, University of Cambridge, United Kingdom

*Abstract*—We fabricated linear arrangements of multiple split-gate devices along an SOI mesa, thus forming a 2×*N* array of individually controllable Si quantum dots (QDs) with nearest neighbor coupling. We implemented two different gate reflectometry-based readout schemes to either probe spin-dependent charge movements by a coupled electrometer with single-shot precision, or directly sense a spin-dependent quantum capacitance. These results bear significance for fast, high-fidelity single-shot readout of large arrays of foundry-compatible Si MOS spin qubits.

## I. Introduction

Quantum computation requires all qubits of a quantum processor to be individually initialized, manipulated, and measured [1]. Among all solid-state implementations, silicon-based spin qubits are especially attractive given the possibility to harness the large-scale integration background of the IC manufacturing industry. In the last few years, significant progress was achieved in this direction with the demonstration of coherent spin control of electrons [2] or holes [3],[4] localized in Si QDs defined by MOS accumulation gates. Yet, establishing a scalable strategy for single-shot sensing of spin states [5] remains one of the key engineering challenges on the path to construct fault-tolerant logical qubits with dynamical error correction codes, including on linear arrays of QDs [6].

## II. Readout in Linear Split-Gate Arrays

The first step towards building a spin qubit register is to localize a well-defined number of elementary charges in a controllable fashion. Field-effect gate electrodes provide a means of shaping the electrostatic landscape in the host Si crystal to form QDs; conventional MOS geometry and process flows naturally lead to linear arrangements of QDs. In our case, lateral confinement perpendicular to the direction of the accumulation gates is provided by mesa patterning. The possibility to couple QDs to their nearest neighbors warrants constraints in terms of gate density. Thus, following a 6nm thermal oxidation of the active area, a gate stack of Poly-Si/5nm TiN was patterned using a hybrid DUV/EBeam process to achieve 64nm pitch. As in a typical CMOS process, charge reservoirs are defined by self-aligned dopant implantation, however the presence of dopant atoms between adjacent QDs is undesirable. Thus, offset spacers larger than half of the gate separation were formed prior to epitaxial source/drain regrowth and n-type dopant implantation (Fig. 1). For large enough channel widths, two pronounced potential minima develop along the upper nanowire edges, leading to distinct QDs sometimes referred to as "corner dots" [7]. Splitting the gates into pairs of face-to-face electrodes enables forming two rows of independently controlled QDs [8]. In the following, we report on cryogenic characterization (T<100mK) of devices featuring multiple pairs of split-gates (Fig. 2), mesa widths W=70-90nm, gate lengths L=40-50nm, lateral and tip-to-tip spacings $S_{side}$=40-50nm and $S_{tip}$=40-60nm.

The aim is to demonstrate the ability to perform single-shot measurements of spin events in such 2×*N* QD arrays, thus laying foundation for a scalable qubit readout method. Though single spins are difficult to sense, it is possible to rely on a spin-charge conversion scheme (Fig. 3), *i.e.* engineer the device geometry and readout sequence so that charge movements become conditioned upon certain spin states. For example, Pauli spin blockade (PSB) can rectify charge transitions between two QDs [9], and energy-selective tunneling into the Fermi sea of a reservoir [10] can help discriminating between excited $|E\rangle$ and ground $|G\rangle$ states. Coupled to a single charge sensing capability, it becomes possible to characterize single spin events. Single-shot charge sensing is typically achieved by probing the time-domain response of an electrometer E (Fig. 4). This electrometer may be a single electron transistor (SET), or any kind of device i/ featuring sharp impedance variations, ii/ having sufficiently strong capacitive coupling to the QD for a single charge event to induce a noticeable impedance shift. One key issue of our prior work on single-shot readout by transport through an embedded SET [11] was that such a configuration could hardly be operational beyond a single split-gate. In the following, we demonstrate two schemes enabling readout in extended linear arrangements, based on gate reflectometry.

## III. Reflectometry on a Coupled Charge Sensor

Time-domain reflectometry consists in monitoring the amplitude and/or phase changes of an RF wave reflecting off a varying load impedance $Z_L$. In this work, an incident carrier signal ($f_{probe}$ several hundreds of MHz) is filtered through a resonant LC circuit connected to one of the gates. Due to a discontinuity of impedance along the transmission line, part of this signal is reflected, with phase/amplitude variations extracted by homodyne detection (Fig. 5a)). A shift in the demodulated signal hence denotes a change of the probed impedance.

The principle of this first scheme is to probe a "single electron box" sensor composed of a QD coupled to a single charge reservoir (or "lead") (Fig. 5b)). In the split-gates array geometry, the electrometer E is strongly coupled to three adjacent $QD_i$ primarily controlled by $V_{Gi}$ (Fig. 5c)). An important aspect of this technique is that sensitivity is good as long as the frequency of charge exchange Γ between the QD receiving the signal and its reservoir is superior to $f_{probe}$ [12]. Yet this tunnel rate Γ may significantly vary with the number of charges confined in the QD, potentially causing signal loss at low occupancy numbers. Therefore, keeping an auxiliary dot E at fixed occupancy (such that $Γ ≥ f_{probe}$) enables sensing the charge transitions occurring in the other QDs down to the last electrons (Fig. 6).

In Fig. 7, the reflectometry signal mapping reveals Coulomb peaks along $V_E$, marking charge transitions within E. Following one of these peaks with decreasing $V_{G1}$ shows several discontinuities, each corresponding to a change in QD1 occupancy – including the unloading of the last electron. The honeycomb pattern typical of double quantum dot (DQD) stability diagrams [13] can be inferred, and QD1 charge domains accurately labeled. While the electrometer potential is primarily controlled by $V_E$, it is also coupled to the neighboring gates. In Fig. 8a), mapping the sensor response in a $V_{G3}(V_{G1})$ plot allows monitoring the charge states in diagonally opposed QDs in the 2×2 array. The same was done in Fig. 8b) between split-gate-defined QD1 and QD2.

Pauli spin blockade can be leveraged to demonstrate the spin readout capability. In our case, thanks to the QDs disposition the improved latched PSB scheme [14],[15] may advantageously be applied (Fig. 9a)). It is based on the fact that by cutting the second QD from the reservoir, the (1,1)→(1,2) transition occurs preferentially via an intermediate (0,2) step, which is not energetically favorable if spins are parallel (*e.g.* |↓↓⟩). Using our sensor, a successful transition to (1,2) should result in a decreased reflectometry signal compared to remaining in (1,1). Fig. 9b) shows that after pulsing the bias from (1,1) to the measurement point (1,2) and following an integration time of 500μs, about half of the traces start with a relatively high readout signal. This is consistent with a charge state stuck in (1,1) and therefore a |↓↓⟩ spin state. After some time, relaxation to |↑↓⟩ occurs (ground state at the measurement point for moderate magnetic fields) and the sensor trace goes down as the charge transition to (1,2) eventually takes place.

Reflectometry on a sensor dot has enabled us to characterize the charge occupancy in the rest of the 2×2 QD array, as well as demonstrate single-shot readout of the spin. However, leads are only available at each end of the line in this particular geometry, which may require sequential shuttling [16] to read a full large-scale array. Next, we implement an alternate readout strategy circumventing this limitation.

## IV. Dispersive readout of quantum capacitance

In this reservoir-free approach, the reflectometry circuit is still connected to the gate of a QD, this time with the goal of sensing a spin-dependent cross-capacitance to an adjacent QD [17]-[19]. For instance, it can be applied to a pair of split-gate-defined dots [20] (Fig. 10). Let the detuning ε be defined as the potential difference between both QDs. The interdot quantum capacitance $C_Q$ is proportional to $\partial^2 E/\partial \varepsilon^2$, *i.e.* the curvature of the eigenstate energy corresponding to the concrete spin configuration vs. detuning. Fig. 11 shows how this curvature varies according to the possible spin eigenstates in a two-electron system. By initializing the spin of the probed "helper" QD2 in |↓⟩, a non-zero $C_Q$ only arises if the spin in QD1 is |↑⟩, in which case the reflected signal undergoes a phase shift.

A $V_{G2}(V_{G1})$ stability diagram is shown on Fig. 12, and plotted for two values of applied static magnetic field. The interdot charge transition (ICT), clearly visible at zero field, disappears at B=2T, which is the hallmark of a blockade. A magnetospectroscopy of this ICT was measured experimentally, showing good agreement with the simulated B-Field response of a singlet/triplet spin blockade [21] (Fig. 13). The asymmetric narrowing of the ICT signal with increasing B corresponds to the increasing ε value at which |↓↓⟩ becomes the ground state. Besides qualitative insight on the nature of the spin blockade, fitting the modeled B-ε behavior (white dashed line in Fig. 13) can provide quantitative information on the lever arm parameter (α, a measure of the gate coupling), minimum separation between states of same spin ($Δ_C$, proportional to tunnel coupling) and electron temperature ($T_e$). The parameters extracted from this ICT were α=0.1eV/V, $Δ_C$=80μeV and $T_e$=270mK.

## V. Conclusion

We demonstrated two gate-based reflectometry readout schemes for probing charge and spin states in linear arrangements of MOS split-gates-defined QDs with leads located at the extremities. Sensing through a lead-coupled electrometer enables precise charge control and single-shot spin readout in all QDs coupled to it. Yet the reach of the sensor is limited in this geometry, which may complicate measurements in large arrays. Complementarily, we showed that dispersive readout of the spin-dependent quantum capacitance between a target and auxiliary QD could be performed far from the reservoirs.

Acknowledgment

This work was supported in part by the EU through H2020 MOS-QUITO and ERC Synergy QuCube projects, and in part by the French Public Authorities through ANR CMOSQSPIN and CODAQ projects.

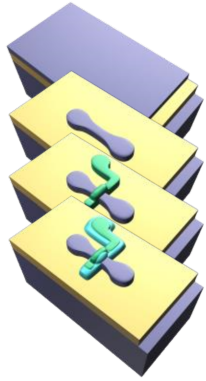

Fig. 1. Simplified process flow of modified SOI mesa devices with 1st offset spacer larger than half of the gate separation.

- 300mm SOI wafers
  $T_{Si}/T_{BOX}$ = 12nm/145nm
- Active mesa patterning
- Thermal oxidation
- MG stack dep. & patterning
  5nm TiN/50nm Poly Si
  64nm min pitch
- 1st spacer
  34nm SiN
- Raised S/D epi
  18nm Si
- LDD implant and anneal
- 2nd spacer
- HDD implant and anneal
- Salicide and BEOL

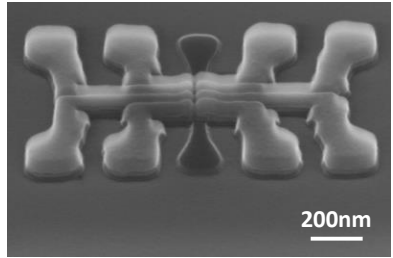

Fig. 2. Tilted SEM view of multiple split-gate pairs (L=40nm) along a mesa (W=70nm) after 1st spacer definition. The spacer (34nm) protects the inter-gate regions ($S_{side}$=$S_{tip}$=40nm) from self-aligned doping, subsequently defining the position of carrier reservoirs at each end of the mesa.

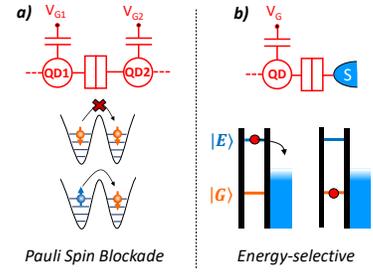

*Spin-charge conversion*

Fig. 3. Two examples of spin-charge conversion: a) Pauli spin blockade between two QDs in series occurring when spins are parallel, b) Energy-selective readout between a QD and a reservoir (S) tuned between "excited" $|E\rangle$ and "ground" $|G\rangle$ spin states.

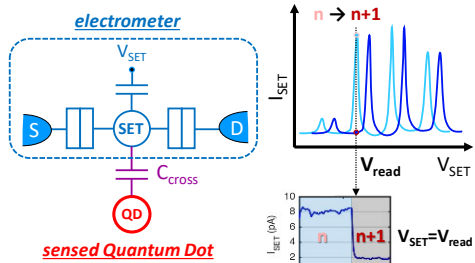

*Single-shot charge sensing*

Fig. 4. Principle of single-shot charge sensing by a capacitively-coupled electrometer, which may be a SET or more generally any device with sharp impedance transitions. A change in electron occupancy of the sensed QD provokes a shift in the transfer characteristics of the sensor. A reading point $V_{read}$ is chosen to maximize the contrast between both charge states. Monitoring the time trace of the sensor provides single-shot charge readout.

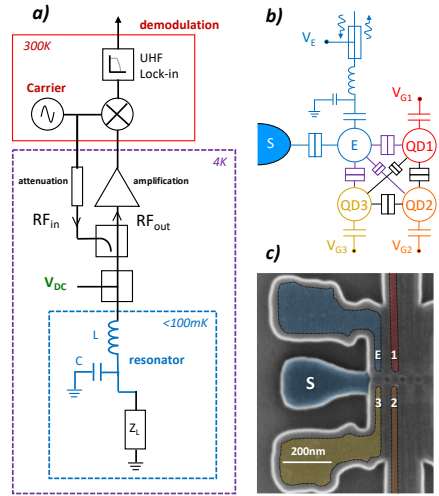

*Reflectometry on a coupled electrometer*

Fig. 5. a) General reflectometry setup. An incident RF signal is sent through an LC resonator to probe changes in a load impedance $Z_L$. A change of $Z_L$ causes a shift in the amplitude and/or phase of the reflected signal. b) The probed impedance in this case is the coupling between an electrometer (E) and a charge reservoir (S). The single electron box E senses charge movements occuring in the neighboring QDs. c) Colored top view SEM of the correspondoing multi-gate device under test.

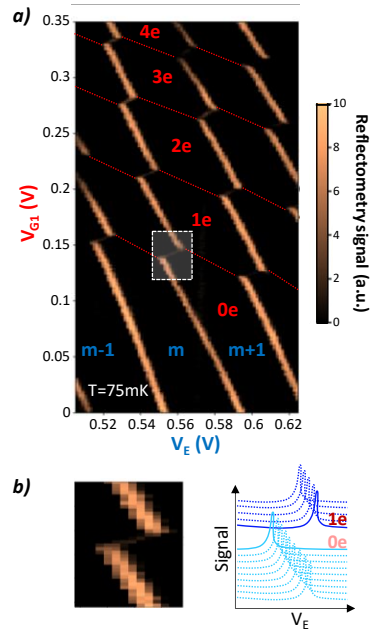

Fig. 6. Benefits of using a remote sensor to detect single electron occupancy in a QD. As the QD charge number decreases, so does its size and outbound tunnel rates (Γ). Sensing through an electrometer of fixed occupancy allows maintaining a favorable ratio between the probing frequency and characteristic time of impedance variation.

Fig. 7. a) 2D colorscale mapping of the reflectometry signal strength in a $V_{G1}$ vs. $V_E$ stability diagram. b) Ridgeline plot illustration of a Coulomb peak steadily drifting with $V_{G1}$ due to cross-talk, until a discontinuity appears due to a charge movement in QD1. A horizontal cutline shows the Coulomb peaks of the electrometer. Following one such Coulomb peak (*e.g.* separating charge domains m and m+1 in E) along the y-axis, clear shifts are observed corresponding to variations of the QD1 occupancy. The signal remains strong down to the unloading of the last electron due to the sensor configuration (Fig. 6). Absolute QD occupancy can thus be assigned to charge domains by counting down the transitions for decreasing values of $V_{G1}$.

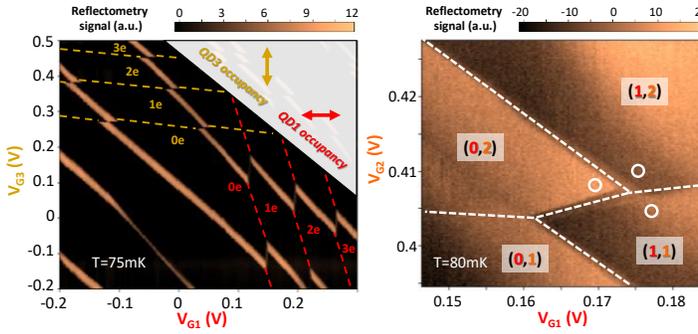
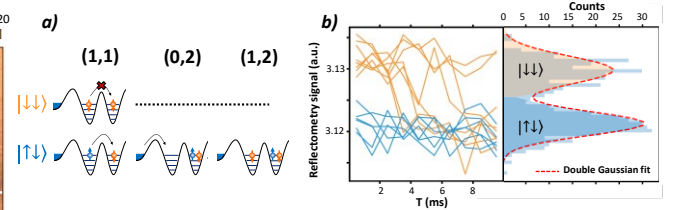

Fig. 8. a) Sensor reflectometry signal for $V_{G3}$ vs. $V_{G1}$, showing the loading of the first electrons in both QD3 and QD1 which are diagonally opposed in the 2x2 array. b) Sensor reflectometry signal for $V_{G2}$ vs. $V_{G1}$, i.e. a split-gate pair. The charge domains of interest in Fig. 9 are marked by white circles. Pauli spin blockade (PSB) may occur between the (1,1) and (0,2) or (1,2) charge states.

Fig. 9. a) Principle of the latched Pauli Spin Blockade (LPSB). Due to geometry, the second QD has much weaker coupling to the reservoir, such that additional electrons are best loaded through the other QD: (1,1)→(0,2)→(1,2). However, this first transition is forbidden if spins are parallel (PSB). Hence, the occurrence of charge movement when pulsing from (1,1) to (1,2) charge domains gives information on the spin. b) Single-shot sensor traces of the spin measurement, and corresponding histogram of measured $|\uparrow\downarrow\rangle$ and $|\downarrow\downarrow\rangle$ states at $T_{integration}$=500µs directly after pulsing to (1,2).

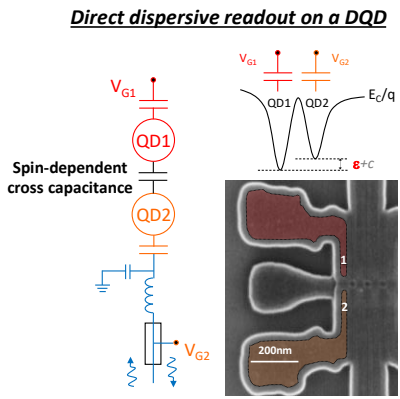
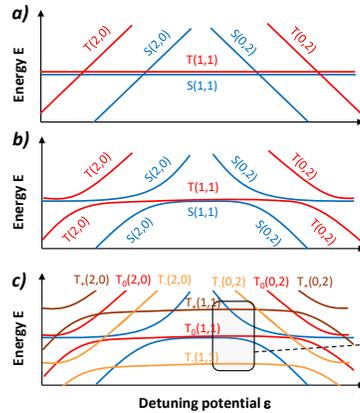
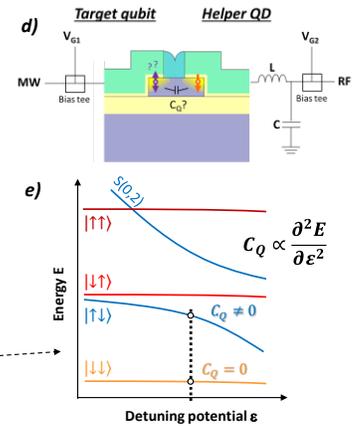

Fig. 10. Double quantum dot direct dispersive readout setup. In this scheme, the carrier reservoirs are not necessary since no charge movement is required to occur. The aim is to probe a spin-dependent quantum capacitance between QD1 and QD2. Detuning potential ε is defined (up to a constant) in relation to the difference between the QD energy minima.

Fig. 11. Principle of spin readout through quantum capacitance sensing in a split-gate device. a) Energy vs. detuning potential ε. When two electrons are in the same QD, the triplet spin states cost extra energy since the second electron has to occupy a higher orbital. b) Turning on the exchange interaction causes singlet states (resp. triplet states) to anticross. c) Applying a magnetic field lifts the degeneracy between the triplet states. d) We consider the (1,1) charge configuration in our device, with helper QD prepared in spin-down. e) We thus focus on the two lowest states. The curved eigenstate $|\uparrow\downarrow\rangle$ gives rise to non-zero $C_Q$, causing a shift of the reflected signal.

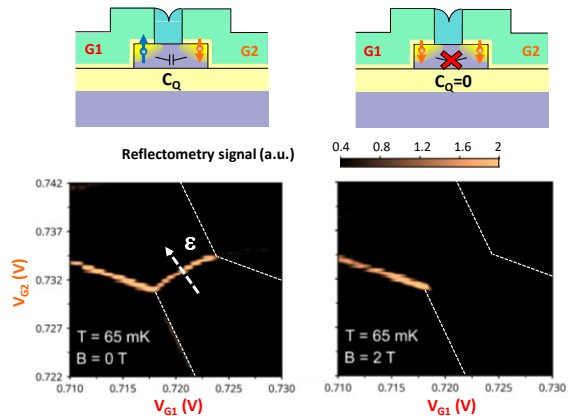
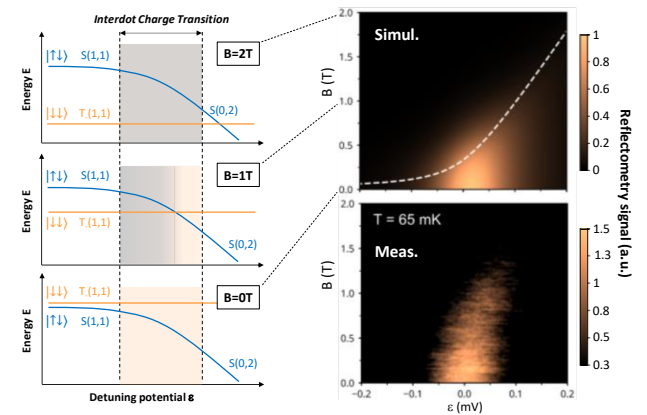

Fig. 12. Double QD stability diagram measured by direct dispersive readout (reflectometry on QD2). Left : at zero DC magnetic field; right : with B = 2T. Applying a B-field results in the disappearance of an interdot charge transition (ICT), which indicates that a blockade occurs. The detuning axis is represented across the ICT.

Fig. 13. Magnetospectroscopy of the ICT on Fig. 12 along the detuning axis, showing the expected behaviour of a Singlet/Triplet spin blockade. As B is increased, the energy of the $|\downarrow\downarrow\rangle$ state is lowered, and the detuning range in which it becomes the ground state expands asymmetrically, leading to the slanted signature reproduced by simulation.